\documentclass[preprintnumbers, prd, twocolumn, showpacs,floatfix,preprintnumbers,
superscriptaddress, nofootinbib]{revtex4}
\usepackage{graphicx}
\usepackage{epsfig}
\usepackage{bm}
\usepackage{amssymb}
\usepackage{float}
\usepackage{amsmath}
\usepackage{subfigure}
\usepackage{dcolumn}
\usepackage{cancel}
\usepackage[colorlinks]{hyperref}
\usepackage[usenames,dvipsnames]{color}
\hypersetup{
     breaklinks=true,
    pdfstartview={FitH},    
    colorlinks=true,       
    linkcolor=blue,          
    citecolor=red,        
    filecolor=magenta,      
    urlcolor=blue,           
    anchorcolor=green,      
    linktocpage=true
}

\def\doi{http://doi.org}

\begin{document}

\title{Study of thermal stability for different dark energy models}
\author{Abdulla Al Mamon}
\email{abdulla.physics@gmail.com}
\affiliation{Department of Mathematics, Jadavpur University, Kolkata-700032, India}
\author{Pritikana Bhandari}
\email{pritikanab@gmail.com}
\affiliation{Department of Mathematics, Jadavpur University, Kolkata-700032, India}
\author{Subenoy Chakraborty}
\email{schakraborty.math@gmail.com}
\affiliation{Department of Mathematics, Jadavpur University, Kolkata-700032, India}
\newcommand{\be}{\begin{equation}}
\newcommand{\ee}{\end{equation}}
\newcommand{\bea}{\begin{eqnarray}}
\newcommand{\eea}{\end{eqnarray}}
\newcommand{\bc}{\begin{center}}
\newcommand{\ec}{\end{center}}
\begin{abstract}
In the present work, we have made an attempt to investigate the dark energy possibility from the thermodynamical point of view. For this purpose, we have studied thermodynamic stability of three popular dark energy models in the framework of an expanding, homogeneous, isotropic and spatially flat FRW Universe filled with dark energy and cold dark matter. The models considered in this work are Chevallier-Polarski-Linder (CPL) model, Generalized Chaplygin Gas (GCG) model and Modified Chaplygin Gas (MCG) model. By considering the cosmic components (dark energy and cold dark matter) as perfect fluid, we have examined the constraints imposed on the total equation of state parameter ($w_{T}$) of the dark fluid by thermodynamics and found that the phantom nature ($w_{T}<-1$) is not thermodynamically stable. Our investigation indicates that the dark fluid models (CPL, GCG and MCG) are thermodynamically stable under some restrictions of the parameters of each model.
\end{abstract}
\maketitle 
\section{Introduction}
During the past two decades, many cosmological observations \cite{Perlmutter:1998np, Riess:1998cb, deBernardis:2000sbo, Percival:2001hw, Spergel:2003cb, Tegmark:2003ud, Komatsu:2010fb} have confirmed that our Universe is undergoing an accelerated expansion phase. In the Einstein theory of gravity, {\it dark energy} (DE), a hypothetical exotic fluid, might be responsible for the presently observed accelerated expansion phase of the Universe (for review, see Ref. \cite{Amendola1}). In this context, various DE models have been proposed to match with observed data and the $\Lambda$CDM ($\Lambda$-Cold-Dark-Matter) model is the most simplest one in this series. Despite the great success of $\Lambda$CDM model, it suffers from two serious theoretical problems, namely the {\it cosmological constant} \cite{Weinberg:2000yb, Padmanabhan:2002ji} and the {\it cosmic coincidence} \cite{Steinhardt:2003st} problems. This clearly motivates people to go deeper into theory for a better understanding of the unknown nature of the DE component.\\
\par As is well known, the thermodynamical study of DE is a powerful apporoach for a better understanding of the unknown nature of the DE component (for details, see Refs. \cite{Barboza:2015rsa,pbshsc17,tspf17,aamss,sdudaam,Lima:2004wf, Callen1,Kubo1}) and thermodynamical aspects are important in this respect. Based on experimental evidences, thermodynamical laws are applicable to different macroscopic systems. However, any thermodynamical process (in contrast to classical mechanics or electromagnetism) does not allow numerical values rather it sets limits on the physical system. So it is expected that the thermodynamic study of the cosmic fluid may unveil the unknown nature of the content of the Universe. Barboza et al. \cite{Barboza:2015rsa} initiated such investigation by analyzing both thermal and mechanical stability of DE fluids. The result of their study are in contradiction with the observational data from the Type Ia Supernova, BAO and Hubble parameter measurements and they concluded that the DE fluid models are not thermodynamically stable. Recently, Bhandari et al. \cite{pbshsc17} have applied the thermodynamical laws and the stability criteria for interacting DE models and there by they are able to impose bounds on the equation of state parameter and constrain the phenomenological interaction term. The present work is an extention of it. Here attempts have been made to examine the thermal stability of three known DE models based on the constraints imposed in \cite{pbshsc17}. The models considered in this work are Chevallier-Polarski-Linder (CPL) model \cite{cpl10,cpl20}, Generalized Chaplygin Gas (GCG) model \cite{cgas1,cgas2,cgas3} and Modified Chaplygin Gas (MCG) model \cite{mcg}. Our study shows that the DE fluid models are thermodynamically stable under some restrictions of model parameters. \\
\par The rest of this paper is organized as follows. In section \ref{sec2}, we have discussed the thermodynamic aspects of different dark energy models in the framework of a spatially flat FRW Universe. Then thermodynamical stability criterion for those models have also been obtained. The discussions and conclusions are summarized in section \ref{sec4}.\\
\par Throughout the paper, we have used the units in which $8\pi G=c=1$.
\section{Thermodynamic aspects of different dark energy models and stability criteria}\label{sec2}
For a homogeneous, isotropic and spatially flat FRW Universe, the Einstein's field equations for the dark fluids, consisting of dark matter (DM) and dark energy (DE), take the following form
\bea 
3H^2=\rho _d+\rho _m \label{eq1}\\
2\dot{H}+3H^2=-p_d-p_m \label{eq2}
\eea
with energy conservation equations
\bea 
\dot{\rho}_{d}+3H(1+w_d)\rho_d=0 \label{eq3}\\
\dot{\rho}_{m}+3H(1+w_m)\rho_m=0\label{eq4}
\eea
where $(\rho _d,p_d, w_d=\frac{p_d}{\rho _d})$  and $(\rho _m, p_m, w_m=\frac{p_m}{\rho _m})$ are the energy density, pressure and the {\it equation of state} (EoS) parameter of the DE and DM components, respectively. Also, $H=\frac{\dot{a}}{a}$ indicates the Hubble parameter, $a(t)$ is scale factor of the Universe and an an overdot $(\cdot)$ denotes differentiation with respect to the cosmic time $t$. Now the above two fluids can be combined as a single fluid having total or effective energy density $\rho _T=\rho_d+\rho_m$, total pressure $p_T=p_d+p_m$ and total EoS parameter $w_T=(w_d \Omega_d + w_m \Omega_m)$. Here $\Omega _d=\frac{\rho_{d}}{3H^2}$ and $\Omega_m=\frac{\rho_{m}}{3H^2}$ are the density parameters of the two dark sectors (DE $\&$ DM). However, in the present work, we have chosen cold dark matter, i.e., $w_m=0$ . As a result, we have $\omega_T=w_d \Omega_d$ and $p_T=p_d$. Thus the equations (\ref{eq1}) and (\ref{eq2}) can effectively be written as
\bea\label{eq5}
3H^2=\rho_T \\
2\dot{H}+3H^2=-p_T
\eea 
Now, the total EoS parameter (with $w_m=0$), in general, can be expressed in terms of $H$ and $a$ as
\be\label{eqwtgen}
w_T=w_d \Omega_d=-1 - \frac{a\frac{dH^2(a)}{da}}{3H^2(a)} 
\ee
\par From the view point of thermodynamic if the expansion of the Universe is considered to be adiabatic and reversible in nature then the energy conservation equation (first law of thermodynamics) takes the following form
\be\label{eq8}
TdS=dI_E+p_T\,dV
\ee
where, $T$, $S$ and $V$ denote the temperature, entropy and volume of the cosmic system respectively. Also, $I_E=V\rho _T$ is the internal energy of the cosmic system and $V=a^3(t)V_0$ is the physical volume of the Universe at a given time and the suffix `0' indicates the value of the corresponding variable at present epoch with $a_0=1$.\\ \\
Also, the fluid equation takes the form 
\be\label{eq9}
d\ln \rho _T + (1+\omega _T) d\ln V=0 
\ee
which is the energy conservation relation. Now, considering $T$ and $V$ as the basic thermodynamical variables, we have obtained (for detailed calculations, see Ref. \cite{pbshsc17})
\be\label{eq12}
I_E=I_0\left(\frac{1+\omega_0}{1+\omega_T}\right)\frac{T}{T_0}
\ee
\par If we consider $V$ as a function of $T$ and $p$, then the variation in $V$ can be obtained as
\be\label{eq19}
dV=V\left(\alpha \,dT-\kappa_T \, dp\right)
\ee
where $\alpha =\frac{1}{V}\left(\frac{\partial V}{\partial T}\right)_p$ and $\kappa _T= -\frac{1}{V}\left(\frac{\partial V}{\partial p}\right)_T$ are known as the thermal expansivity and the isothermal compressibility respectively. Similarly, the adiabatic compressibility ($\kappa _s$) is defined by keeping $S$ to be fixed as
\be\label{eq22}
\kappa _s= -\frac{1}{V}\left(\frac{\partial V}{\partial p}\right)_S
\ee
The relation among compressibilities and heat capacities in isothermal and adiabatic thermodynamical scenarios is given by \cite{Kubo1}
\be\label{eq24}
\frac{\kappa _T}{\kappa _s}=\frac{C_p}{C_v}
\ee
where, $C_p$ and $C_v$ are termed as the heat capacity at constant pressure and the heat capacity at constant volume respectively.\\
\par  The above thermodynamical parameters are related by (for details, see Refs. \cite{Barboza:2015rsa,pbshsc17})
\bea
C_p=\left(\frac{\partial (I_E+ pV)}{\partial T}\right)_p \nonumber ~~~~~~~~~~~~~~~~~~~~~~~~~~~ \\
=\frac{(1+\omega _T)I_E}{T} =\frac{(1+\omega _0)I_0}{T_0}= \mbox{constant}
\eea
\be
C_v=\left(\frac{\partial I_E}{\partial T}\right)_V=\frac{C_p ~d\ln V}{\{(1+w_T)d\ln V-d\ln w _T\}}
\ee
\be
\alpha =\frac{C_p}{V\rho_T(1+\omega_T)}\left[\frac{d\omega_T}{\omega_T \{d\omega_T -\omega_T (1+\omega_T) d\ln V\}}+1\right]
\ee
\be
\kappa _s=\frac{d\ln V}{\{(1+\omega_T) d\ln V-d\ln \omega_T\}}\kappa _T
\ee
\be
\kappa _T=\frac{\alpha V}{C_p}
\ee 
Thus, we can use the above relations to impose bounds on the EoS parameter of the cosmic fluid. We shall now discuss the thermal stability of the present system assuming work done only due to the volume variation of the thermal system. For stable equilibrium, the second order variation of the internal energy becomes, $\delta ^2 I_E=\delta T\,\delta S-\delta p\,\delta V \geq 0$ \cite{Barboza:2015rsa, Kubo1} . Now choosing $(S, p)$ or $(T, V)$ as the independent thermodynamical variables, the variation term $``\delta ^2 I_E"$ can be obtained as
$$\delta ^2 I_E =V \kappa _s \delta p^2 + \frac{T}{C_p} \delta S^2 ~~{\rm or}~~\delta ^2 I_E =\frac{1}{V \kappa _T} \delta V^2 +\frac{C_v}{T} \delta T^2$$
Therefore, for stability of the system, the thermodynamical parameters $(C_p, C_v, \kappa _s, \kappa _T)$ should be non negative (i.e., $C_p, C_v, \kappa_s, \kappa_T \geq 0$). The non-negativity of the thermodynamical parameters are studied from the above interrelations and the corresponding results are given in table \ref{table1}.
\begin{center}
	\begin{table*}[ht]
		\renewcommand{\arraystretch}{2}
		\caption{Conditions for Thermodynamical stability criteria. The corresponding expressions for $w_T$ are given in equations (\ref{eqwtacpl}), (\ref{eqwtagcg}) and (\ref{eqwtamcg}) for the CPL, GCG and MCG models, respectively.} \label{table1}
		\begin{tabular}{| >{\centering\arraybackslash}m{3cm}|>{\centering\arraybackslash}m{5cm}|}
			\hline
			{\bf Restriction on} \bm{$w_T$} & {\bf Condition for stability} \\
			\hline
			$w_T <-1$ & unstable \\
			\hline
			$w_T >0$ &  $\frac{dw_T}{da}<\frac{3w _T ^2}{a} ~~\mbox{or}~~ \frac{dw_T}{da} > \frac{3w_T (1+w_T)}{a}$ \\
			\hline
			$-1< w_T <0$ & $\frac{3w_T (1+w_T)}{a}<\frac{dw_T}{da} < \frac{3w_T ^2}{a}$\\
			\hline
		\end{tabular}
	\end{table*}
\end{center}
\par In the following, we shall consider three different DE models and for each one we shall study the criterion for thermodynamical stability.
\subsection{Chevallier-Polarski-Linder (CPL) model} 
Firstly, we have considered CPL parametrization for the DE EoS parameter \cite{cpl10,cpl20} which has been widely used by all the recent cosmological observations to put constrain on the cosmological parameters. In CPL model, the EoS parameter is parametrized as  
\be\label{cpl}
w_{d}(a) = w_{0} + w_{1}(1 - a) 
\ee
where $w_{0}$  represents the current value of $w_{d}(a)$ and $w_{1}$ accounts for the variation of the EoS parameter with respect to the scale factor $a$. From the infinite past ($a = 0$) till the present epoch ($a = 1$), the EoS parameter varies between $w_0 +w_1$ and $w_0$. \\
The solution for $\rho_{d}$ from equation (\ref{eq3}) is obtained as
\be
\rho_{d}(a) = \rho_{d0} a^{-3(1 + \omega_{0} + \omega_{1})} e^{-{3w_{1}(1-a)}}
\ee
where, $\rho_{d0}$ is the present value of $\rho_{d}(a)$. Similarly, the solution for $\rho_{m}$ from equation (\ref{eq4}) is obtained as
\be
\rho_{m}(a) = \rho_{m0} a^{-3}
\ee
where, $\rho_{m0}$ is the present value of $\rho_{m}(a)$.\\
For this model, from equation (\ref{eq1}), the Hubble expansion rate can be obtained as
\be\label{eqhacpl}
H^2(z) = H^2_{0}\left[\Omega_{m0} a^{-3} + \Omega_{d0} a^{-3(1 + w_{0} + w_{1})} e^{-{3w_{1}(1-a)}}\right]
\ee 
with $\Omega_{m0}+\Omega_{d0}=1$. Here, $H_{0}$ is the Hubble parameter at the present epoch, $\Omega_{m0} = \frac{\rho_{m0}}{3H^2_{0}}$ and $\Omega_{d0} = \frac{\rho_{d0}}{3H^2_{0}}$ are the density parameters at the present epoch of the cold DM and DE components respectively.\\
In this case, the expression for $w_T$ is obtained, using equations (\ref{eqwtgen}) and (\ref{eqhacpl}), as
\be\label{eqwtacpl}
w_{T}(a)=\frac{w_0 -2 +w_1 (1-3a)}{3 + \frac{3\Omega_{m0}}{1-\Omega_{m0}} a^{(w_0 + w_1 -2)}e^{{3w_{1}(1-a)}}}
\ee
\begin{figure}[ht]
\begin{center}
\includegraphics[width=0.27\textwidth,height=0.20\textheight]{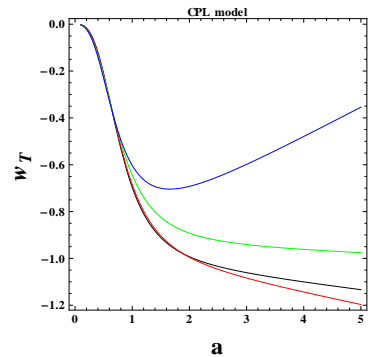}
\caption{The evolution of $w_{T}$ is shown for CPL model by taking $\Omega_{m0}=0.315$ from \cite{pom0} and different values of the parameter pair  ($w_{0}$, $w_{1}$). The black, red, green and blue curves are for ($-0.99$, $0.03$), ($-0.9$, $0.05$), ($-0.8$, $0.01$) and ($-0.9$, $-0.13$) respectively.}
\label{figwtcpl}
\end{center}
\end{figure}
\begin{figure}[ht]
\begin{center}
\includegraphics[width=0.27\textwidth,height=0.20\textheight]{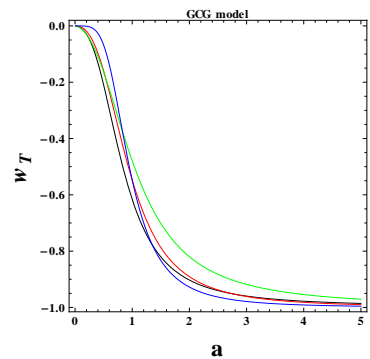}
\caption{The evolution of $w_{T}$ is shown for GCG model taking $\Omega_{m0}=0.315$ \cite{pom0} and different values of the parameter pair  ($A_{s}$, $\alpha$). The black, red, green and blue curves are for ($0.9$, $-0.5$), ($0.8$, $-0.2$), ($0.7$, $-0.4$) and ($0.8$, $0.5$) respectively.}
\label{figwtgcg}
\end{center}
\end{figure}
\begin{figure}[ht]
\begin{center}
\includegraphics[width=0.27\textwidth,height=0.20\textheight]{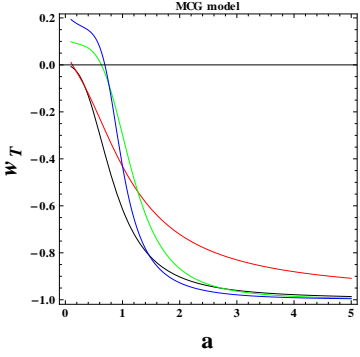}
\caption{The evolution of $w_{T}$ is shown for MCG model taking $\Omega_{m0}=0.315$ \cite{pom0} and different values of the model parameters ($B_{s}$, $\alpha$, $\gamma$). The black, red, green and blue curves are for ($0.9$, $-0.5$, $0.03$), ($0.7$, $-0.7$, $0.23$), ($0.5$, $0.3$, $0.13$) and ($0.7$, $0.7$, $0.23$) respectively.}
\label{figwtmcg}
\end{center}
\end{figure}
\begin{figure}[ht]
\begin{center}
\includegraphics[width=0.32\textwidth,height=0.20\textheight]{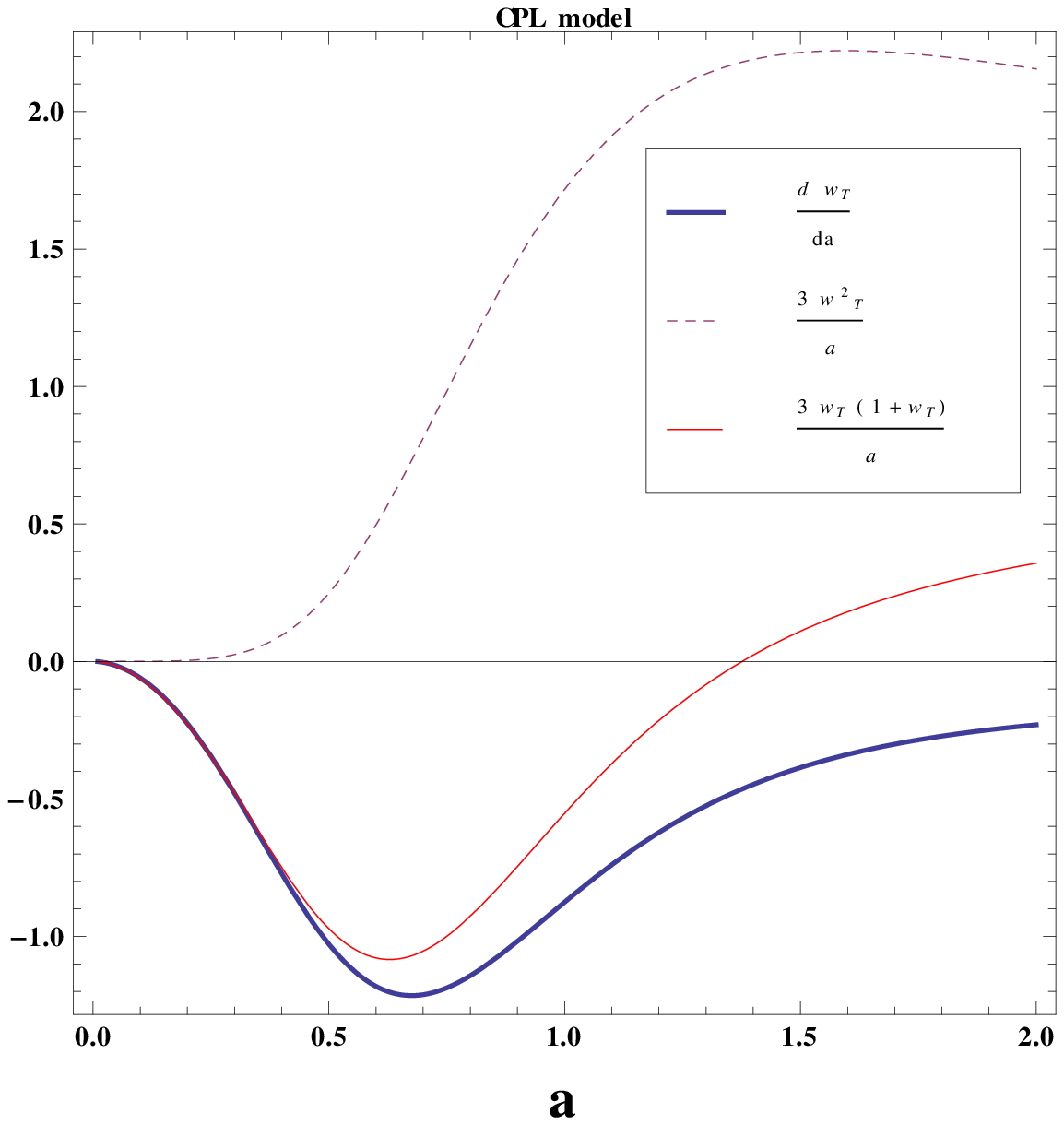}\\
\includegraphics[width=0.32\textwidth,height=0.20\textheight]{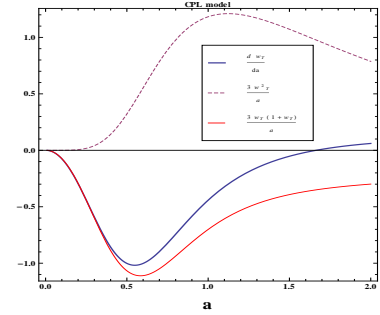}
\caption{Stability of the CPL model is shown by taking $\Omega_{m0}=0.315$ \cite{pom0} and different values of the $w_{0}$ and $w_{1}$. The upper panel is for $w_0=-0.99$ and $w_1=0.15$, while the lower panel corresponds to $w_0=-0.99$ and $w_1=-0.13$.}
\label{figscpl}
\end{center}
\end{figure}
\subsection{Generalized Chaplygin Gas (GCG) model} 
It is well known that the energy density of the GCG behaves as a cold DM in the past and as a cosmological constant at present \cite{cgas1,cgas2,cgas3}. Motivated by this idea, in this paper, we are interested to explain the late-time evolution of the Universe produced by the GCG. For this purpose, we have considered that the Universe contains both the cold DM and the GCG. The EoS parameter of the GCG is given by \cite{cgas1,cgas2,cgas3}
\be\label{eqgcg}
p_{d} = - \frac{A}{\rho^{\alpha}_{d}},~~~~~0< \alpha \leq 1
\ee
where $A>0$. The case $\alpha = 1$ corresponds to the original chaplygin gas first introduced by Chaplygin \cite{cgas1}. Using equations (\ref{eq3}) and (\ref{eqgcg}), one finds that the energy density of the GCG evolves as
\be\label{rhogcg}
\rho_{d}(a) = {\left[A + Ba^{-3(1+\alpha)}\right]}^{\frac{1}{(1+\alpha)}}
\ee
where, $B$ is an integration constant. Equation (\ref{rhogcg}) can be re-written in the following form
\be
\rho_{d}(a) = \rho_{d0}{\left[A_{s} + (1-A_{s})a^{-3(1+\alpha)}\right]}^{\frac{1}{(1+\alpha)}}
\ee
where, $A_{s}=\frac{A}{A+B}$ and $\rho_{d0} = {\left(A + B\right)}^\frac{1}{(1+\alpha)}$ is the current value of the energy density of the GCG. It deserves mention here that equation (\ref{eqgcg}) will represent a generalized polytropic gas if $\alpha<0$. To ensure the finite and positive value of $\rho_{d}$ we require $-1< \alpha \leq 1$ and $0\le A_{s}\leq 1$. \\ 
\par The Hubble parameter is now given by
\be
H^2 = H^2_{0}{\left[ \Omega_{m0}a^{-3} + \Omega_{d0}{\left(A_{s} + (1-A_{s})a^{-3(1+\alpha)}\right)}^{\frac{1}{(1+\alpha)}}\right]}
\ee
It is obvious that the GCG will behave like cosmological constant $\Lambda$ when we put $A_{s}=1$.\\
For this model, $w_T$ evolves as
\be\label{eqwtagcg}
w_{T}(a)= \frac{A_s a^{3(1+\alpha)}f_{GCG}(a)}{1-A_{s}(1-a^{3(1+\alpha)}){(-\Omega_{m0}-f_{GCG}(a))}}
\ee
where, $$f_{GCG}(a)=(1-\Omega_{m0})a^{3}{\Big[ A_s + (1-A_s)a^{-3(1+\alpha)} \Big]}^{\frac{1}{1+\alpha}}$$.
\begin{figure}[ht]
\begin{center}
\includegraphics[width=0.32\textwidth,height=0.20\textheight]{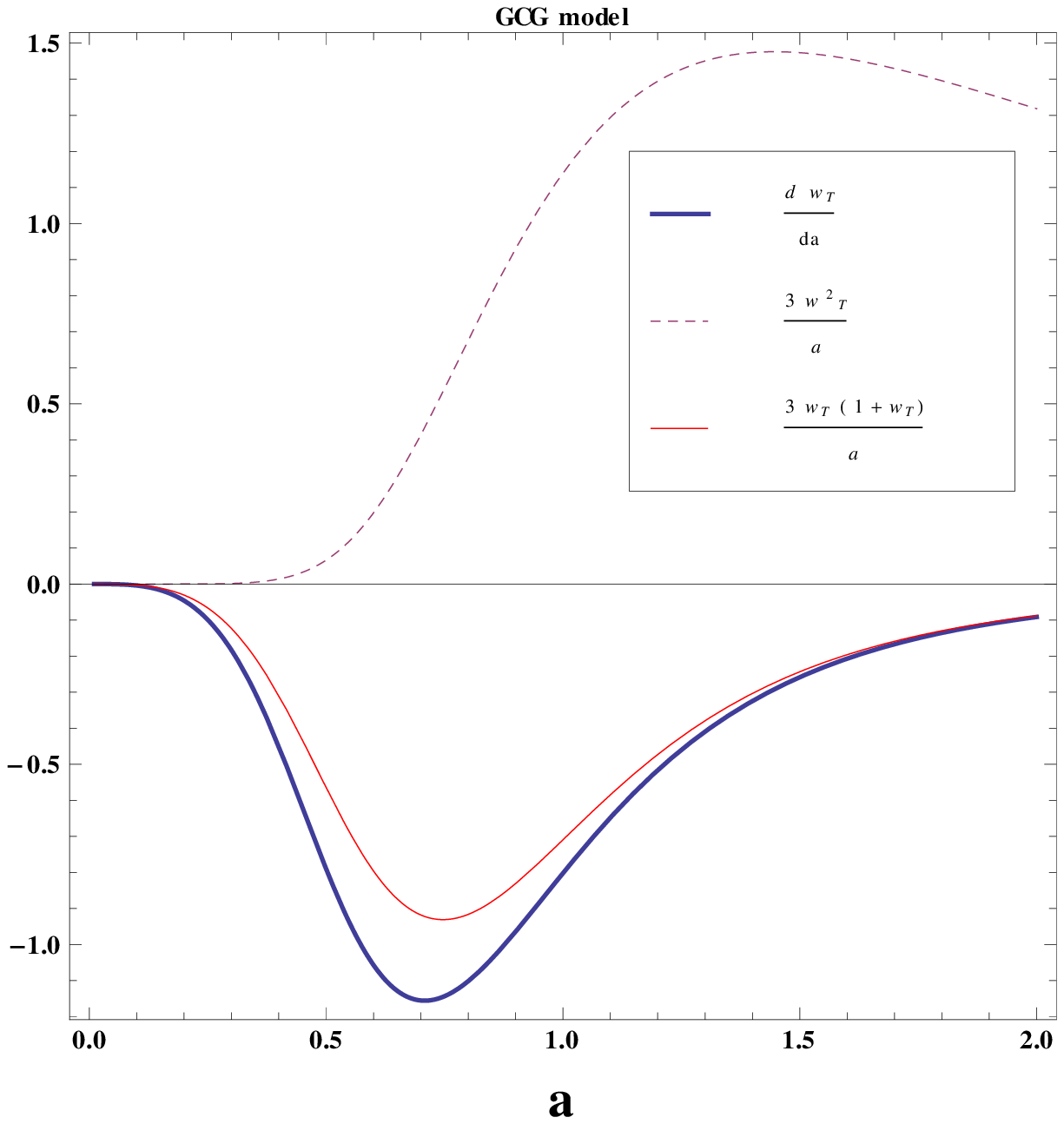}\\
\includegraphics[width=0.32\textwidth,height=0.20\textheight]{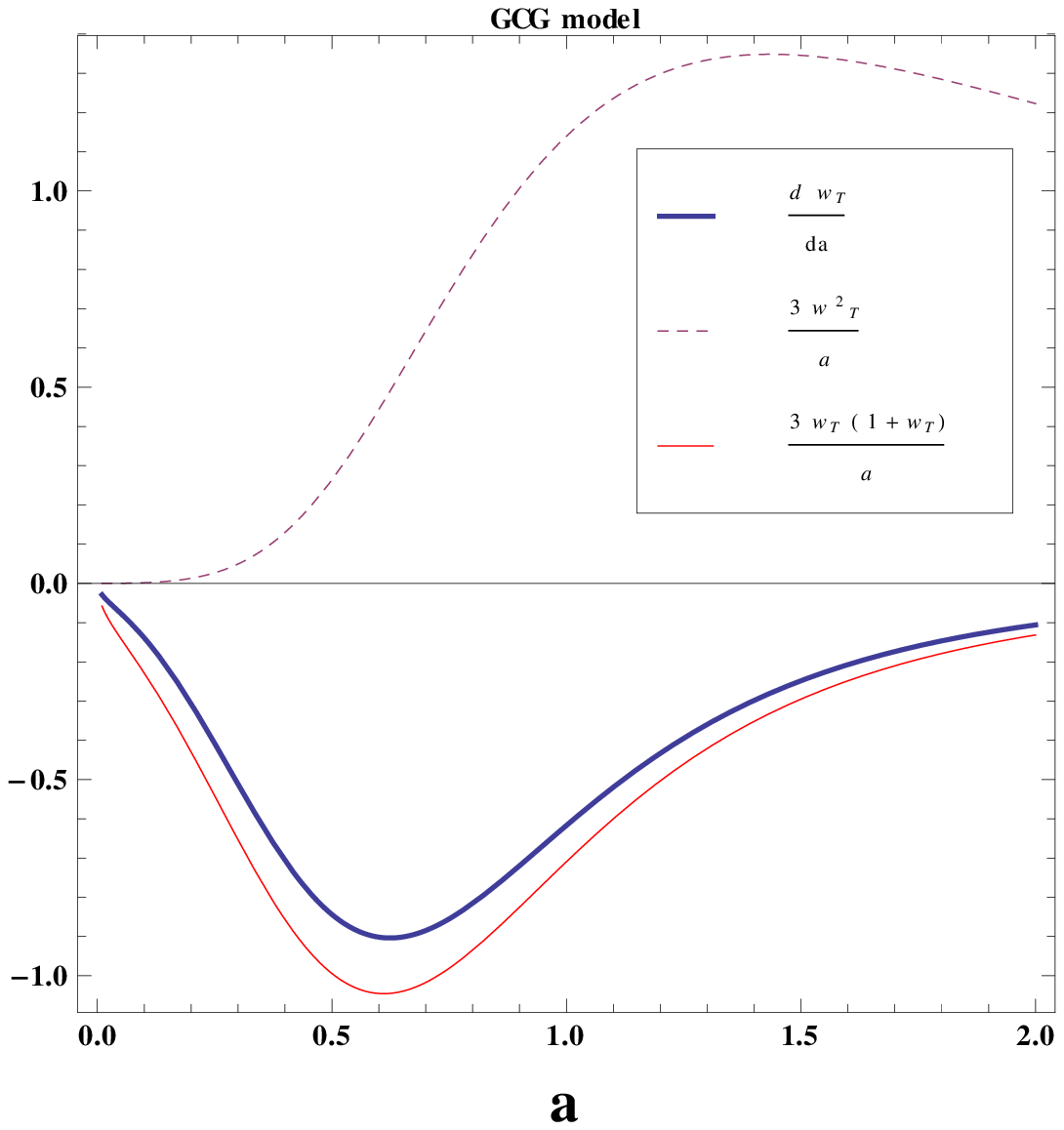}
\caption{Stability of the GCG model is shown by taking $\Omega_{m0}=0.315$ \cite{pom0} and different values of the $A_{s}$ and $\alpha$. The upper panel is for $A_s=0.9$ and $\alpha=0.5$, while the lower panel corresponds to $A_s=0.9$ and $\alpha=-0.5$.}
\label{figsgcg}
\end{center}
\end{figure}
\subsection{Modified Chaplygin Gas (MCG) model}
Next, we have considered that the MCG plays the role of DE. The MCG is characterized by the following EoS parameter \cite{mcg}
\be\label{eqmcg}
p_{d}=\gamma \rho_{d} - \frac{A}{\rho^{\alpha}_{d}},~~~~~0<\gamma\leq 1
\ee
where $\gamma$, $A$ and $\alpha$ are model parameters. It is clear that the GCG model, as given in equation (\ref{eqgcg}), is recovered when the parameter $\gamma$ is zero. On the other hand, if $A = 0$, the MCG model looks like a perfect fluid with constant EoS parameter $w = \gamma$. Also, the MCG model mimics a cosmological constant when $\alpha = -1$ and $A = 1 + \gamma$. Like the GCG, equation (\ref{eqmcg}) will represent a modified polytropic gas if $\alpha<0$. From Equation (\ref{eqmcg}), one can find that $0 \leq B_{s} \leq 1$ and $-1< \alpha \leq 1$ are required to keep the finite and positivity of the MCG energy density. The energy density for MCG takes the form
\be
\rho_{d}(a)=\rho_{d0}{\left[B_{s} + (1-B_{s})a^{-3(1+\gamma)(1+\alpha)}\right]}^{\frac{1}{(1+\alpha)}},~~~\gamma\neq -1
\ee
where, $B_{S}=\frac{A}{(1+\gamma)\rho^{(1+\alpha)}_{d0}}$. In this case, the Hubble parameter is given by
\be
H^2 = H^2_{0}{\left[ \Omega_{m0}a^{-3} +   
\Omega_{d0}{\left(B_{s} + (1-B_{s})a^{-3(1+\gamma)(1+\alpha)}\right)}^{\frac{1}{(1+\alpha)}}\right]}
\ee
In this case, $w_T$ evolves as
\bea\label{eqwtamcg}
w_{T}(a)= \frac{f_{MCG}(a)}{1-B_{s}(1-a^{3(1+\alpha)(1+\gamma)})}~~~~~~~~~~~~~~~\nonumber \\
\times \frac{B_s a^{3(1+\alpha)(1+\gamma)}- \gamma(1-B_s)}{-\Omega_{m0}-f_{MCG}(a)}
\eea
where, $$f_{MCG}(a)=(1-\Omega_{m0})a^{3}{\Big[ B_s + (1-B_s)a^{-3(1+\alpha)(1+\gamma)} \Big]}^{\frac{1}{1+\alpha}}$$.
\begin{figure}[ht]
\begin{center}
\includegraphics[width=0.32\textwidth,height=0.20\textheight]{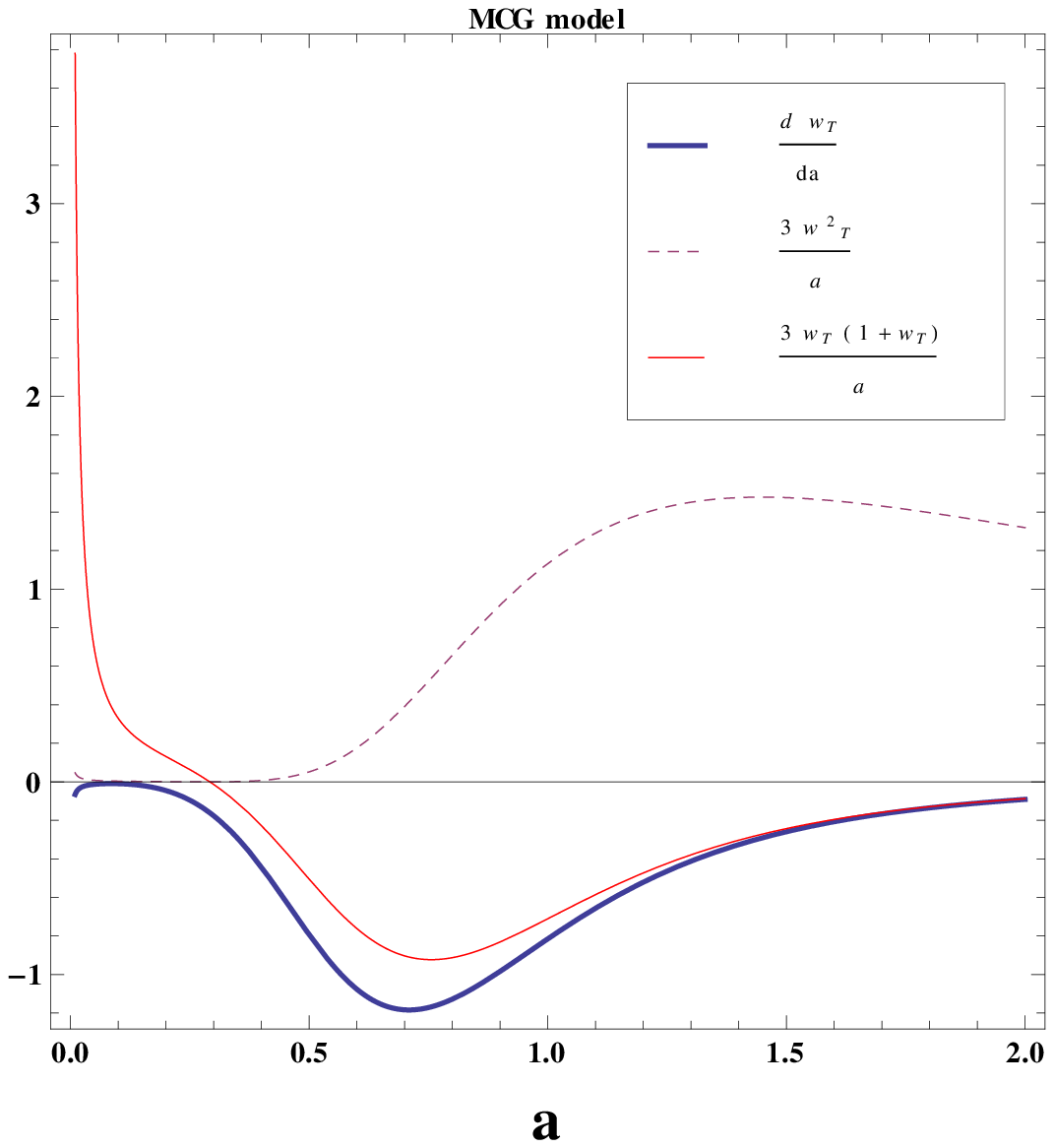}\\
\includegraphics[width=0.32\textwidth,height=0.20\textheight]{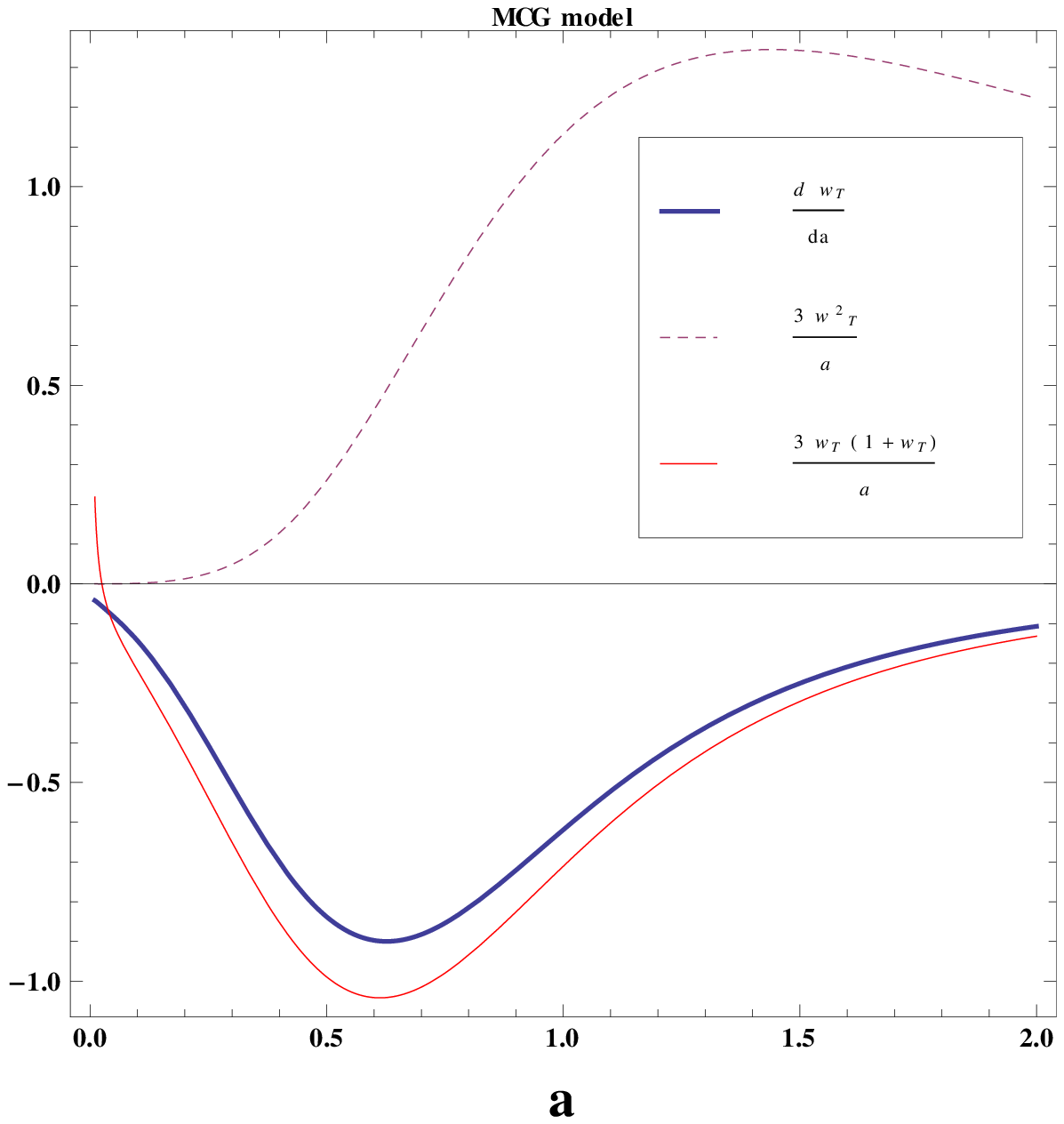}
\caption{Stability of the MCG model is shown by taking $\Omega_{m0}=0.315$ \cite{pom0} and different values of the $B_{s}$, $\alpha$ and $\gamma$. The upper panel is for $B_s=0.9$, $\alpha=0.5$ and $\gamma=0.03$, while the lower panel corresponds to $B_s=0.9$, $\alpha=-0.5$ and $\gamma=0.03$.}
\label{figsmcg}
\end{center}
\end{figure}
\section{Discussion and Concluding Remarks} \label{sec4}
As mentioned before, in the present work, we have investigated the thermodynamical aspects of three popular dark energy models (namely CPL, GCG and MCG models) in the framework of an expanding, homogeneous, isotropic and spatially flat FRW Universe filled with dark fluids (cold DM and DE). As it is expected that the dark fluid should reach the thermodynamic stability, so such a requirement has been shown in table \ref{table1} for a general $w_T$. It is also evident from table \ref{table1} that the phantom nature ($w_{T}<-1$, leading singularity problems) of the cosmic fluid is not thermodynamically stable. In figure \ref{figwtcpl}, \ref{figwtgcg} and \ref{figwtmcg}, we have shown the evolution of $w_T$ by assuming different values of the model parameters for the CPL, GCG and MCG models, respectively. Using the thermodynamical stability criteria, given in table \ref{table1}, we have also examined the stability graphically for each model. For CPL model, we have shown the thermodynamical stability for different values of the parameter pair $(w_{0},w_{1})$ in figure \ref{figscpl}. It has been found that for the CPL model, the stability conditions hold for negative values of $w_1$ (as $w_0$ is always negative). Similarly, for GCG and MCG models, we have also shown the thermodynamical stability for different values of the parameter pairs ($(A_{s},\alpha)$ for GCG model and $(B_{s},\alpha,\gamma)$ for GCG model) in figure \ref{figsgcg} and \ref{figsmcg}, respectively. It has been found that the stability conditions hold only for negative values of $\alpha$ for each model. Thus the thermodynamical stability conditions always hold for the generalized/modified polytropic gas models ($\alpha<0$). We therefore conclude that the aforesaid models are thermodynamically stable under some restrictions of model parameters. 
\section{Acknowledgments}
AAM acknowledges the financial support from Science and Engineering Research Board (SERB), Govt. of India through its National Post-Doctoral Fellowship Scheme (File No: PDF/2017/000308). PB acknowledges DST-INSPIRE for awarding Research fellowship. SC is thankful to IUCAA, Pune, India for research facilities at Library. SC also acknowledges the UGC-DRS Programme in the Department of Mathematics, Jadavpur University.

\end{document}